\documentclass[letterpaper, 10 pt, conference]{ieeeconf}  
\IEEEoverridecommandlockouts                              
\usepackage{times}
\usepackage{xcolor}
\usepackage[pdftex]{graphicx}
\usepackage{hyperref}
\usepackage[small]{caption}
\usepackage{subcaption}

\usepackage{amsmath,amsfonts,amssymb}
\usepackage{prettyref}
\usepackage{mathrsfs}
\usepackage{graphicx}
\usepackage{wrapfig}
\usepackage{makecell}
\usepackage{multirow}
\usepackage{booktabs}
\usepackage{algorithm}
\usepackage{algpseudocode}
\algnewcommand{\LineComment}[1]{\State \(\triangleright\) #1}
\algdef{SE}[DOWHILE]{Do}{doWhile}{\algorithmicdo}[1]{\algorithmicwhile\ #1}
\newcommand*{\colorboxed}{}
\def\colorboxed#1#{%
  \colorboxedAux{#1}%
}
\newcommand*{\colorboxedAux}[3]{%
  \begingroup
    \colorlet{cb@saved}{.}%
    \color#1{#2}%
    \boxed{%
      \color{cb@saved}%
      #3%
    }%
  \endgroup
}

\newrefformat{Fig}{Fig.~\ref{#1}}
\newrefformat{fig}{Fig.~\ref{#1}}
\newrefformat{par}{Sec.~\ref{#1}}
\newrefformat{appen}{Appendix~\ref{#1}}
\newrefformat{sec}{Sec.~\ref{#1}}
\newrefformat{sub}{Sec.~\ref{#1}}
\newrefformat{table}{Table~\ref{#1}}
\newrefformat{tab}{Table~\ref{#1}}
\newrefformat{alg}{Algorithm~\ref{#1}}
\newrefformat{Alg}{Algorithm~\ref{#1}}
\newrefformat{Def}{Definition~\ref{#1}}
\newrefformat{Thm}{Theorem~\ref{#1}}
\newrefformat{step}{Step~\ref{#1}}
\newrefformat{ln}{Line~\ref{#1}}
\newrefformat{eq}{Eq.~\ref{#1}}
\newrefformat{pb}{Problem~\ref{#1}}
\newrefformat{it}{Item~\ref{#1}}
\newrefformat{te}{Term~\ref{#1}}
\def\Eqref Eq:#1:{\eqref{eq:#1}}
\newrefformat{Eq}{Equation~\Eqref#1:}

\newcommand{\E}[1]{\mathbf{#1}}

\newcommand{\pref}[1]{\prettyref{#1}}

\DeclareMathOperator{\arrival}{arrival}
\DeclareMathOperator{\collision}{collision}
\DeclareMathOperator{\smooth}{smooth}

\DeclareMathOperator{\rob}{rob}
\DeclareMathOperator{\obs}{obs}

\title{\large\bf DeepMNavigate: Deep Reinforced Multi-Robot Navigation Unifying Local \& Global Collision Avoidance
\vspace{-15px}
}
\author{Qingyang Tan$^{1}$, Tingxiang Fan$^{2}$, Jia Pan$^{2}$, Dinesh Manocha$^{1}$\\
Video Link: \url{https://youtu.be/LWLBxWuwPeU}
\vspace{-60px}
\thanks{$^1$Qingyang Tan and Dinesh Manocha are with Department of Computer Science and Electrical \& Computer Engineering, University of Maryland at College Park. \{qytan,dm@cs.umd.edu\} $^2$Tingxiang Fan and Jia Pan are with Department of Computer Science, University of Hong Kong. \{tingxfan@hku.hk, jpan@cs.hku.hk\}}
\thanks{Project website: https://qytan.com/publication/global\_planning/}
}
        
\begin{document}
\maketitle
\thispagestyle{empty}
\pagestyle{empty}

\begin{abstract}
We present a novel algorithm (DeepMNavigate) for global multi-agent navigation in dense scenarios using deep reinforcement learning (DRL). Our approach uses local and global information for each robot from motion information maps. We use a three-layer CNN that takes these maps as input to generate a suitable action to drive each robot to its goal position. Our approach is general, learns an optimal policy using a multi-scenario, multi-state training algorithm,  and can directly handle raw sensor measurements for local observations. We demonstrate the performance on dense, complex benchmarks with narrow passages and environments with tens of agents. We highlight the algorithm's benefits over prior learning methods and geometric decentralized algorithms in complex scenarios.
\end{abstract}



\section{Introduction}\label{sec:intro}

Multi-robot systems are increasingly being used for different applications, including surveillance, quality control systems, autonomous guided vehicles, warehouses, cleaning machines, etc. A key challenge is to develop efficient algorithms for navigating such robots in complex scenarios while avoiding collisions with each other and the obstacles in the environment. As larger numbers of robots are used, more efficient methods are needed that can handle dense and complex scenarios.

\begin{figure}[h]
\centering
\includegraphics[width=0.35\textwidth]{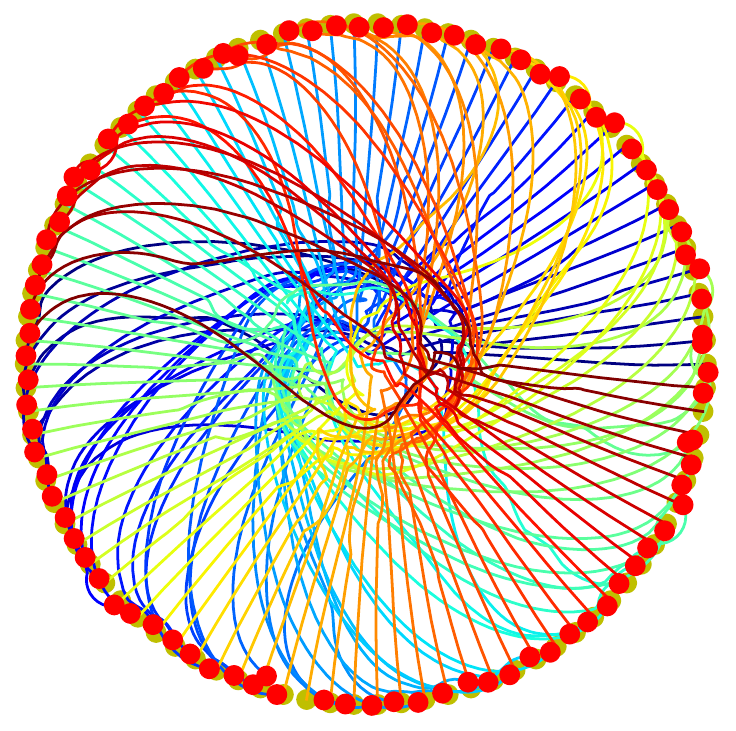}
\caption{{\bf Circle Crossing:} Simulated trajectories of $90$ robots in circle crossing scenarios generated by our algorithm that uses global information. The yellow points are the initial positions of the robots and the red points are the diametrically opposite goals of the robot. Our DeepMNavigate algorithm can handle such scenarios without collisions along the trajectories, and all the robots reach their goals. Prior learning methods that only use local methods~\cite{long2018towards,fan2018fully,JHow1} cannot handle such scenarios, as the robots tend to get stuck.}
\label{fig:traj_circle}
\vspace{-10px}
\end{figure}

Multi-agent navigation has been studied extensively in robotics, AI, and computer animation. At a broad level, previous approaches can be classified into centralized~\cite{van2005prioritized,van2009centralized,luna2011push,sanchez2002using,solovey2016hardness,yu2018effective} or decentralized planners~\cite{van2011reciprocal,geraerts2008using,helbing1995social,fox1997dynamic}. 
One benefit of decentralized methods is that they can scale to a large number of agents,  though it is difficult to provide any global guarantees on the resulting trajectories~\cite{fraichard2004inevitable} or to handle challenging scenarios with narrow passages (see Figures \ref{fig:traj_circle} or \ref{fig:corridor}). 

 Recently, there has been considerable work on developing new, learning-based planning algorithms~\cite{fan2018fully,CesarCadena,JHow1,JHow2,long2018towards,long2017deep} for navigating one or more robots through dense scenarios. Most of these learning methods  tend to learn an optimal policy using a multi-scenario, multi-stage training algorithm. 
 However, current learning-based methods are limited to using only the local information and do not exploit any global information. Therefore, it is difficult to use them in dense environments or narrow passages.

\noindent{\bf Main Results:} We present a novel, multi-robot navigation algorithm (DeepMNavigate) based on reinforcement learning that exploits {\em a combination of local and global information}. We use a multi-stage training scheme that uses various multi-robot simulation scenarios with global information. In terms of training, we represent the global information using a {\em motion information} map that includes the location of each agent or robot. We place the robot information in the corresponding position to generate a bit-map and use the resulting map as an input to a three-layer CNN. Our CNN considers this global information along with local observations in the scene to generate a suitable action to drive each robot to the goal without collisions. We have evaluated our algorithm in dense environments with many tens of robots (e.g., 90 robots) navigating in tight scenarios with narrow passages.  
As compared to prior multi-robot methods, our approach offers the following benefits:
\begin{enumerate}
     \item We use global knowledge in terms of motion information maps in our network to improve the performance of DRL. This also results in higher reward value. 
     \item Our approach scales with the number of robots and is able to compute collision-free and smooth trajectories. Running our trained system takes many tens of seconds on a PC with a 32-core CPU and one NVIDIA RTX 2080 Ti on multi-robot systems with $10-90$ robots.
     \item We can easily handle challenging multi-robot scenarios like inter-changing robot positions or multiple narrow corridors, which are difficult for prior geometric decentralized or local learning methods.  In particular, we highlight the performance on five difficult environments that are very different from our training scenarios and have more agents. This demonstrates the generalizability of our method.
\end{enumerate}

\section{Related Work}

\subsection{Geometric Multi-Robot Navigation Algorithms}

Most prior algorithms are based on geometric techniques such as sampling-based methods, geometric optimization, or complete motion planning algorithms. The centralized methods assume that each robot has access to complete information about the state of the other robots based on some global data structure or communication system~\cite{luna2011efficient, sharon2015conflict, yu2016optimal, tang2018hold} and compute a safe, optimal, and complete solution for navigation. However, they do not scale to large multi-robot systems with tens of robots. Many pratical geometric decentralized methods for multi-agent systems are based on reciprocal velocity obstacles~\cite{van2011reciprocal} or its variants~\cite{alonso2013optimal}. 
These synthetic  methods can be used during the training phase of learning algorithms.

\subsection{Learning-Based Navigation Methods}

Learning-based collision avoidance techniques usually try to optimize a parameterized policy using the data collected from  different tasks. 
Many navigation algorithms adopt a supervised learning paradigm to train collision avoidance policies. 
Muller et al.~\cite{muller2006off}  present a  vision-based static obstacle avoidance system using a 6-layer CNN to map input images to steering angles. Zhang et al.~\cite{zhang2017deep} describe a successor-feature-based deep reinforcement learning algorithm for robot navigation tasks based on raw sensory data. Barreto et al.~\cite{barreto2017successor} apply transfer learning to deploy a policy for new problem instances. Sergeant et al.~\cite{sergeant2015multimodal} propose an approach based on multimodal deep autoencoders that enables a robot to learn how to navigate by observing a dataset of sensor inputs and motor commands collected while being tele-operated by a human. Ross et al.~\cite{ross2013learning} adapt an imitation learning technique to train reactive heading policies based on the knowledge of a human pilot. Pfeiffer et al. \cite{pfeiffer2017perception} map the laser scan and goal positions to motion commands using expert demonstrations. To be effective, these methods need to collect training data in different environments, and the performance is limited by the quality of the training sets.

To overcome the limitations of supervised-learning, Tai et al. \cite{tai2017virtual} present a mapless motion planner trained end-to-end without any manually designed features or prior demonstrations. Kahn et al. \cite{kahn2017uncertainty} propose an uncertainty-aware model-based learning algorithm that estimates the probability of collision, then uses that information to minimize the collisions at training time. To extend learning-based methods to highly dynamic environments, some decentralized techniques have been proposed. Godoy et al. \cite{godoy2016moving} propose a Bayesian inference approach that computes a plan that minimizes the number of collisions while driving the robot to its goal. Chen et al. \cite{JHow1,chen2017socially} and Everett et al. \cite{JHow2} present multi-robot collision avoidance policies based on deep reinforcement learning, requiring the deployment of multiple sensors to estimate the state of nearby agents and moving obstacles. Yoon et al. \cite{yoon2019learning} extend the framework of centralized training with decentralized execution to perform additional optimization for inter-agent communication. Fan et al. \cite{fan2018fully} and  Long et al. \cite{long2018towards, long2017deep}  describe a decentralized multi-robot collision avoidance framework where each robot makes navigation decisions independently without any communication with other agents. It has been extended in terms of multiple sensors and explicit pedestrian motion prediction~\cite{sathyamoorthy2020densecavoid}. Other methods account for social constraints~\cite{JHow2}.
However, all these methods do not utilize global information about the robot or the environment, which could be used to improve the optimality of the resulting paths or handle challenging narrow scenarios. 

\begin{figure}[!htb]
\captionsetup[subfigure]{justification=centering}
\centering
\begin{subfigure}{0.35\textwidth}
\includegraphics[width=1\textwidth]{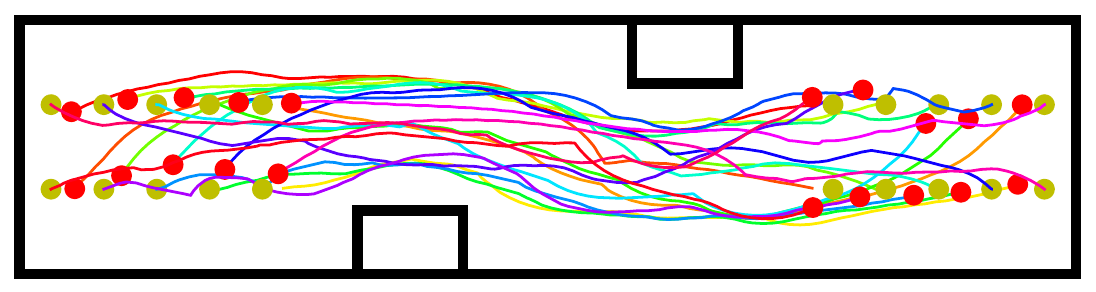}
\caption{\footnotesize{Our algorithm is able to compute collision-free trajectories for $20$ robots.}}
\label{fig:corridor_traj}
\end{subfigure}
\begin{subfigure}{0.35\textwidth}
\includegraphics[width=1\textwidth]{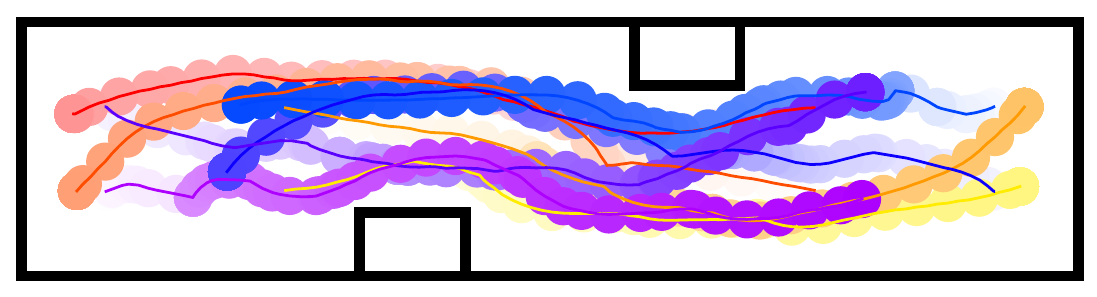}
\caption{\footnotesize{We highlight some of the computed trajectories with temporal information computed using our algorithm.}}
\label{fig:corridor_traj_temporal}
\end{subfigure} 
\begin{subfigure}{0.35\textwidth}
\includegraphics[width=1\textwidth]{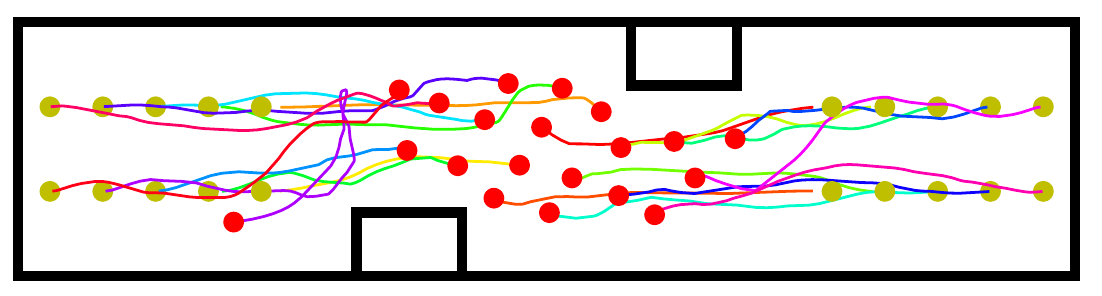}
\caption{\footnotesize{Trajectories by local learning method~\cite{long2018towards}. All agents do not reach the goal position.}}
\label{fig:ori_corridor_traj}
\end{subfigure}
\begin{subfigure}{0.35\textwidth}
\includegraphics[width=1\textwidth]{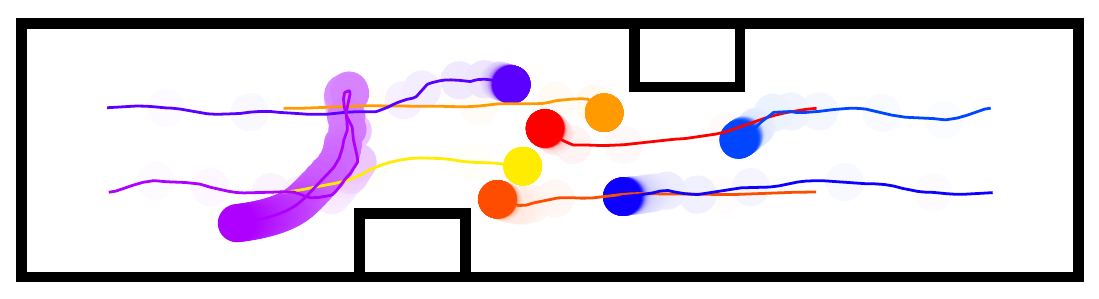}
\caption{\footnotesize{Selected trajectories by \cite{long2018towards} with temporal information. The agents get stuck.}}
\label{fig:ori_corridor_traj_temporal}
\end{subfigure}
\caption{{\bf Narrow Corridor:}
We compute the trajectories computed by DeepMNavigate and~\cite{long2018towards} for two groups of robots ($20$ total) exchanging their positions through  narrow corridors. In (a) and (c), the yellow points correspond to the initial positions and the red points correspond the final positions. (b) and (d), highlight the temporal information along the trajectories using color and transparency.   Prior local planning methods~\cite{long2018towards} can only handle these scenarios with up to 12 agents and the  geometric decentralized methods~\cite{van2011reciprocal} cannot handle such cases. This benchmark is quite different from training datasets.}
\label{fig:corridor}
\vspace{-10px}
\end{figure}

\section{Multi-Robot Navigation}

\subsection{Problem Formulation and Notation}\label{formulation}

We consider the multi-robot navigation problem for non-holonomic differential drive robots. Our goal is to design a scheme that avoids collisions with obstacles and other robots and works well in dense and general environments. We describe the approach for 2D, but it can be extended to 3D workspaces and to robots with other dynamics constraints.

Let the number of robots be $N_{\rob}$. We represent each robot as a disc with radius $R$. At each timestep $t$, the $i$-th robot $(1 \leq i \leq N_{\rob})$ has access to an observation $\E{o}_{i}^{t}$ and then computes an action $\E{a}_{i}^{t}$ that drives the $i$-th robot towards its goal $\E{g}_{i}^{t}$ from the current position $\E{p}_{i}^{t}$. The observation of each robot includes four parts: $\E{o}^{t} = [\E{o}_{z}^{t} ,\E{o}_{g}^{t} , \E{o}_{v}^{t}, \E{o}_{M}^{t}]$, where $\E{o}_{z}^{t}$ denotes the sensor measurement (e.g., laser sensor) of its surrounding environment, $\E{o}_{g}^{t}$ stands for its relative goal position, $\E{o}_{v}^{t}$ refers to its current velocity, and $\E{o}_{M}^{t}$ is the robot motion information, which includes the global state of the system, discussed in Section \ref{sec:global}. In this paper, we focus on analyzing and incorporating motion information in the navigation system.  Meanwhile, there are $N_{\obs}$ static obstacles in the environment. We use $\E{B}_{k}$ to denote the area occupied by a static $k$-th obstacle. The computed action $\E{a}^{t}$ drives the robot to its goal while avoiding collisions with other robots and obstacles within the timestep $\Delta t$ until the next observation $\E{o}^{t+1}$ is received.

Let $\mathbb{L}$ be the set of trajectories for all robots, subject to the robot's kinematic constraints, i.e.:
{\small
\begin{align}
    & \mathbb{L}=\{ l_i, i=1,...,N_{rob} | 
    \E{v}_{i}^{t}\sim\pi_\theta(\E{a}_{i}^{t}|\E{o}_{i}^{t}),
     \E{p}_{i}^{t}=\E{p}_{i}^{t-1}+\Delta t\cdot \E{v}_{i}^{t}, \nonumber \\
    &\forall j \in [1,N_{\rob}],j \neq i, 
    \left\| \E{p}_{i}^{t} - \E{p}_{j}^{t} \right\| > 2R \wedge
    \forall k \in [1,N_{\obs}], \nonumber \\
    & \forall \E{q} \in \E{B}_k, 
    \left\| \E{p}_{i}^{t} - \E{q} \right\| > R \wedge
    \left\|\E{v}_{i}^{t}\right\| \leq v_{i}^{\max}  \},
\end{align}
}
where $p_i^{0}$ is the initial position of the robot and $p_i^t$ are the positions at timestep $t$. $v_i^t$ is the current linear velocity on the 2D plane (i.e. $v_x, v_y$) as a result of the action $a_i^t$. The agent we are simulating is non-holonomic and can only control the linear velocity on  the $X$ axis and the angular velocity on the $Z$ axis (which is used to describe the rotation in the 2D plane).

\subsection{Multi-Agent Navigation Using Reinforcement Learning}\label{sec:fan_work}
Our approach builds on prior reinforcement learning approaches that use local information comprised of various observations. Some of them  only utilize three of the four elements mentioned in \ref{formulation}. The term $\E{o}_{z}^{t}$ may include the measurements of the last three consecutive frames from a sensor. 
The relative goal position $\E{o}_{g}^{t}$ in these cases is a 2D vector representing the goal position in polar coordinates with respect to the robot’s current position. The observed velocity $\E{o}_{v}^{t}$ includes the current velocity of the robot. These observations are normalized using the statistics aggregated during training\cite{wiki:mean_std, implementation_ppo}. This normalization can  make RL training more stable and improve its performance. The action of a differential robot includes the translational and rotational velocity, i.e. $\E{a}^{t}=[v^t,\omega^t], v \in (0,1), \omega \in (-1,1)$.
We use the following reward function to guide a team of robots:
\begin{align}
    r_{i}^{t}=(^{g}r)_{i}^{t}+(^{c}r)_{i}^{t}+(^{\omega}r)_{i}^{t}.
\end{align}
When the robot gets closer or reaches its goal, it is rewarded as
{\footnotesize
\begin{align}
(^{g}r)_{i}^{t}=
\begin{cases}
r_{\arrival} & \text{if } \left\| \E{p}_{i}^{t} - \E{g}_{i} \right\| < 0.1\\
r_{\text{approaching}}(\left\| \E{p}_{i}^{t-1} - \E{g}_{i} \right\|-\left\| \E{p}_{i}^{t} - \E{g}_{i} \right\|)& \text{otherwise.}
\end{cases}
\end{align}
}
The $\|\E{p}_{i}- \E{g}_{i} \|$ item denotes the distance between the robot and its goal. When there is a collision, it is penalized using the function
{\small
\begin{align}
(^{c}r)_{i}^{t}=
\begin{cases}
r_{\collision} & \text{if } \left\| \E{p}_{i}^{t} - \E{p}_{j}^{t} \right\| < 2R \\
& \text{or } \left\| \E{p}_{i}^{t} - \E{q} \right\| < R, \E{q} \in \E{B}_k\\
0& \text{otherwise.}
\end{cases}
\end{align}
}
In addition to collision avoidance, one of our goals is generating a smooth path. A simple technique is to impose penalties whenever there are large rotational velocities. Although it is not a standard way to obtain smooth path, we found this technique can achieve a smooth trajectory empirically. This can be expressed as
{\small
\begin{align}
(^{\omega}r)_{i}^{t}=
\begin{cases}
    r_{\smooth}|\omega_{i}^t| & \text{if } |\omega_{i}^t|>0.7. \\
    0& \text{otherwise,}
\end{cases}
\end{align}
}
where $r_{\arrival}$, $r_{\text{approaching}}$, $r_{\collision}$ and $r_{\smooth}$ are parameters used to control the reward. These parameters provide reward feedback for the agents, which makes the training process more stable~\cite{nair2018overcoming}. In practice, the reward parameters can be tuned to obtain desirable behaviors (e.g., learn more conservative behaviors without a larger collision penalty). We do not change the reward function when we use our approach for different environments.

\begin{figure}[t!]
    \centering
    \includegraphics[width=0.45\textwidth]{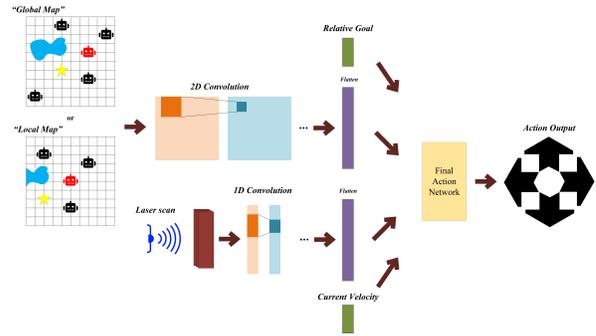}
    \caption{We highlight the architecture of our policy network (DeepMNavigate), including  ({\em global} and {\em local}) maps used by our approach. The {\em global map} is based on the world coordinate system and each {\em local map} is centered at the corresponding robot's current location. The red robot represents the map's corresponding robot, black robots represent the neighboring robots, the yellow star represents the goal, and the blue area is an obstacle. In our implementation, the map is discretized and assigned  different values.  We use a 2D convolutional neural network to handle the additional global information input from the map and fully-connected network to compute the action for each robot.}
    \label{fig:pipeline}
    \vspace{-15px}
\end{figure}


\section{DeepMNavigate: Trajectory Computation Using Global Information}\label{sec:global}

In this section, we present our novel, learning-based, multi-agent navigation algorithm that uses positional information of other agents. Our formulation is based on motion information maps and uses a $3$-layer CNN to generate a suitable action for each agent.

\subsection{Motion Information Maps} \label{input}

Prior rule-based decentralized methods such as \cite{van2011reciprocal} use information corresponding to the position and velocity of each agent to compute a locally-optimal and collision-free trajectory.  Our goal is to compute similar state information to design better learning-based navigation algorithms.  
Such state information can be either gathered based on some communication with nearby robots or computed using a deep network that uses raw sensor data.
In our formulation, we use maps that consist of each agent's location as input. 
In particular, we use two different map representations: one corresponds to all the robots based on the world coordinate system  and is called the {\em global-map};  the other map is centered at each robot's current location and uses the relative coordinate system and is  called the {\em local-map}.

We use the following method to compute the global-map and the local-map. During each timestep $t$, we specify the $i$-th robot's position in the world frame as $\E{x}^t_i\in\mathbb{R}^2$. We also use the goal positions $\E{g}_i, \forall 1\leq i\leq N_{\rob}$, obstacle information $\E{B}_k, \forall 1\leq k\leq N_{\obs}$, to build the map $M^t_i\in\mathbb{R}^{h\times w}$ for the $i$-th robot. Assume the size of the simulated robot scenario is $H\times W$, where $H$ represents the height, $W$ represents the width, and the origin of the world frame is located at $(\frac{H}{2}, \frac{W}{2})$. Each pixel of the map ($M_i^t(p, q), \forall 1\leq p \leq h, 1\leq q \leq w$) indicates which kind of object lies in the small area $\mathcal{A}_{pq} = \left(\frac{(p-1)H}{h}-\frac H2, \frac{pH}{h}-\frac H2\right]\times\left(\frac{(q-1)W}{w}-\frac W2, \frac{qW}{w}-\frac W2\right]$ in the world frame. Assuming each object's radius is $r_i$, then $M_i^t(p,q)$ corresponds to:

{
\small
\begin{align}
M_i^t(p,q) = \left\{
\begin{array}{ll}
1,& \{y|\|y-\E{x}^t_i\|\leq R\}\bigcup\mathcal{A}_{pq}\neq \emptyset\\
2& \exists 1\leq j\leq N_{\rob},\ j\neq i ,\\
&\quad s.t.\ \{y|\|y-\E{x}^t_j\|\leq R\}\bigcup\mathcal{A}_{pq}\neq \emptyset\\
3,& \{y|\|y-\E{g}_i\|\leq R\}\bigcup\mathcal{A}_{pq}\neq \emptyset\\
4,& \exists 1\leq k\leq N_{\obs}, s.t.\ \E{B}_k\bigcup\mathcal{A}_{pq}\neq \emptyset\\
0,& \text{otherwise},
\end{array}
\right..
\label{eq:map}
\end{align}
}
where ``1'' represents the corresponding robot, ``2'' represents the neighboring robots, ``3'' represents the robot's goal, ``4'' represents the obstacles and ``0'' represents  the empty background (i.e. free space). This is highlighted in \pref{fig:pipeline}.

In some scenarios, there could be a restriction on the robot's movement in terms of static obstacles or regions that are not accessible. Our global-map computation takes this into account in terms of representing $M_i^t(p,q)$. However, these maps may not capture scenes with no clear boundaries or very large environments spread over a large area.  If we use a {\em global-map} with the world coordinate representation for all the agents, the resulting map would be extremely large and would involve a high computational cost and memory overhead. 
In these cases, we use the {\em local-map} for each agent, instead of considering the whole scenario, which whold have a size $H\times W$  These local maps only account for information in a relatively small neighborhood with a fixed size $H_l\times W_l$.
The size of the local neighborhood ($H_l,\ W_l$) can be tuned to find a better performance for different applications. In addition to the position information, these maps may contain other state information of the robot, including velocity, orientation, or dynamics constraints as additional channels.

\subsection{Proximal Policy Optimization}

We use proximal policy optimization \cite{schulman2017proximal} to optimize the overall system. This training algorithm has the stability and reliability of trust-region methods: i.e., it tries to compute an update at each step that minimizes the cost function while ensuring that the deviation from the previous policy is relatively small. The resulting proximal policy algorithm updates the network using all the steps after several continuous simulations (i.e. after each robot in the entire system reaches its goal or stops running due to collisions) instead of using only one step to ensure the stability of network optimization. In these cases, if we store the robot positions or motion information as dense matrices corresponding to the formulation in \pref{eq:map}, it would require a significant amount of memory and also increase the overall training time. Instead, we use a sparse matrix representation for $M_t^i$. We compute the non-zero entries of$M_t^i$ based on the current position of each robot, the goal positions and obstacle information using \pref{eq:map}. To feed the input to our neural network, we generate dense representations using temporary sparse storage. This design choice allows us to train the system using trajectories from 58 agents performing 450 actions using only 2.5GB memory. More details on the training step are given in \pref{sec:exp}.

\subsection{Network}

To analyze a large matrix and produce a low-dimensional feature for input $M^t_i$, we use a convolutional neural network (CNN) because such network structures are useful for handling an image-like input (or our map representation). Our network has three convolutional layers with architecture, as shown in \pref{fig:pipeline}.  Our network extracts the related locations of different objects in the environment and could guide the overall planner to avoid other agents and obstacles to reach the goal. 

Our approach to handling raw sensor data (e.g., 2D laser scanner data) uses the same structure as local methods\cite{long2018towards}, i.e. a two-layer, 1D convolutional network. Overall, we use a two-layer, fully-connected network that takes as input the observation features, including the feature generated by the 1D \& 2D CNNs, related goal position, and observed velocity. This generates the action output and local path for each robot. We highlight the whole pipeline of the network for local and global information in \pref{fig:pipeline} and \pref{alg:policy}.

\begin{algorithm}
    \caption{Policy Making for DeepMNavigate }
    \label{alg:policy}
    \begin{algorithmic}[1]
   
    \For {timestep $t = 1, 2,... $}
      \State // \textit{Each robot runs individually in parallel}
      \For{$\text{robot } i = 1, 2, ... N$}
      \State Collect observation $\E{o}_{i}^{t}=[\E{o}_{z}^{t} ,\E{o}_{g}^{t} , \E{o}_{v}^{t}, \E{o}_{M}^{t}]$
      \State {\footnotesize Run policy $\pi_{\theta}$ represented by the network, get action $\E{a}_i^t$}
      \State {\small Update the robot position $\E{p}_i^t$ according to action $\E{a}_i^t$}
      \EndFor
    \EndFor
\end{algorithmic}
\end{algorithm}

\begin{table}[!htbp]
    \vspace{-15pt}
    \centering
    \setlength{\tabcolsep}{4pt}
    \renewcommand{\arraystretch}{1.5}
    \scalebox{0.6}{
    \begin{tabular}{|c|c|c|c|c|c|}
        \hline
        Layer & Convolutional Filter & Stride & Padding & Activation Fuction& Output Size \\
        \hline 
        Input & - & - & - & - & $250\times250\times1$ \\

        \hline
        Conv 1 & $7 \times 7 \times 1 \times 8$ & $1\times 1$ & `SAME' &ReLU& $250\times250\times8$\\
        \hline
        Max Pooling 1 & $3 \times 3$ & $2\times 2$ & `SAME' &- & $125\times125\times8$\\

        \hline
        Conv 2 & $7 \times 7 \times 8 \times 12$ & $1\times 1$ & `SAME' &ReLU& $125\times125\times12$\\
        \hline
        Max Pooling 2 & $3 \times 3$ & $2\times 2$ & `SAME'&-  & $63\times63\times12$\\

        \hline
        Conv 3 & $7 \times 7 \times 12 \times 20$ & $1\times 1$ & `SAME' &ReLU& $63\times63\times20$\\
        \hline
        Max Pooling 3 & $3 \times 3$ & $2\times 2$  & `SAME'&- & $32\times32\times20$\\

        \hline
        Flatten & - & - & - & - & $20480$\\
        \hline
        Fully connected & $20480\times128$& - & - & ReLU & $384$\\
        \hline
        Fully connected & $128\times64$& - & - & ReLU & $256$\\
        \hline
    \end{tabular}
    }
    \caption{Architecture and hyper-parameters of a convolutional neural network that considers the global information.}
    \label{tab:global_network}
    \vspace{-10pt}
\end{table}

\begin{figure}[ht]
\centering
\includegraphics[width=0.45\textwidth]{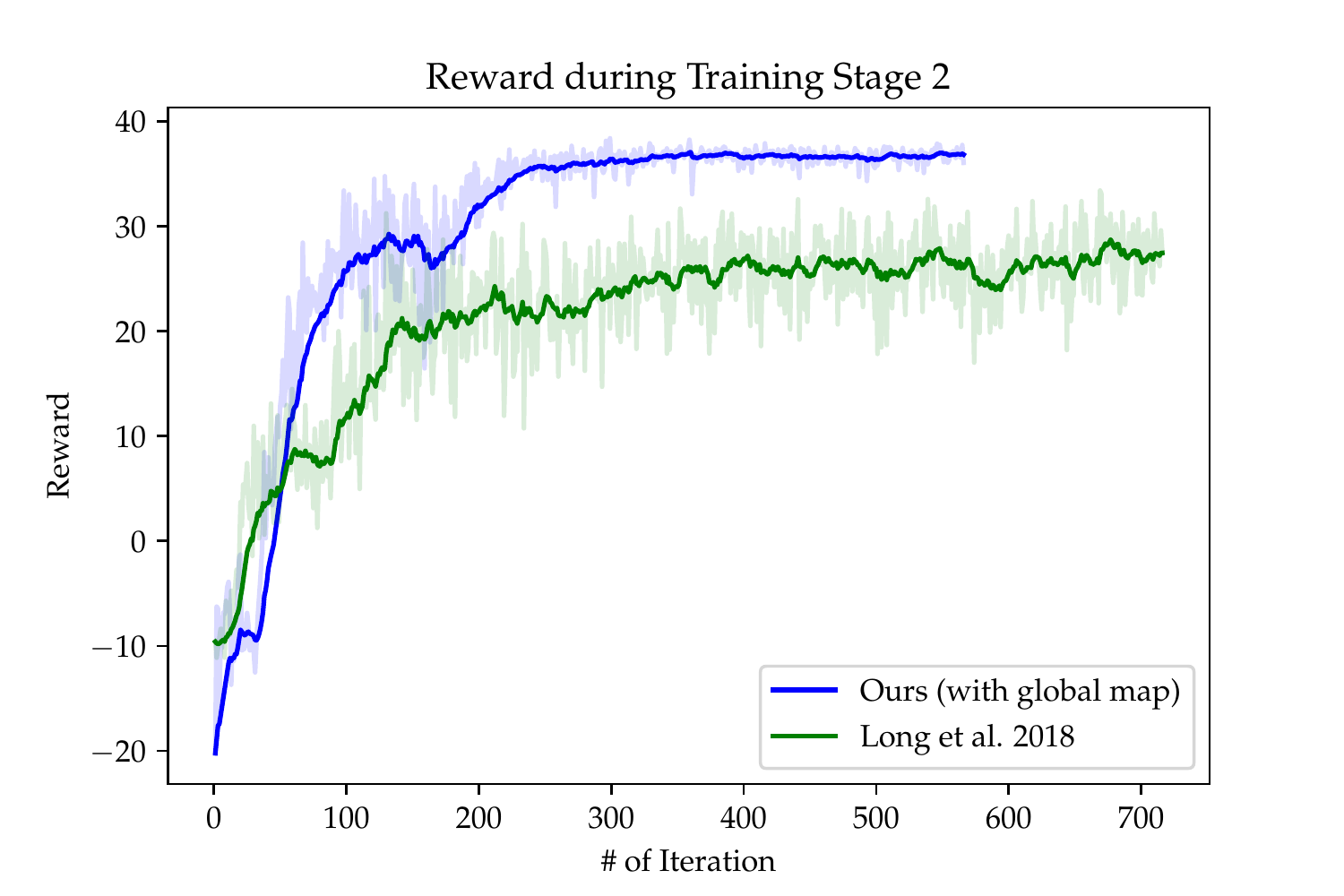}
\vspace{-8px}
\caption{We highlight the reward as a function of the number of iterations during the second stage of the overall training algorithm. We compare the performance of a reinforcement learning algorithm   that only uses local information~\cite{long2018towards} to our method, which uses local and global information. Our method obtains a higher reward than~\cite{long2018towards} due to the global information. }
\label{fig:reward}
\vspace{-12px}
\end{figure}

\begin{figure}[h]
    \centering
    \includegraphics[width=0.45\textwidth]{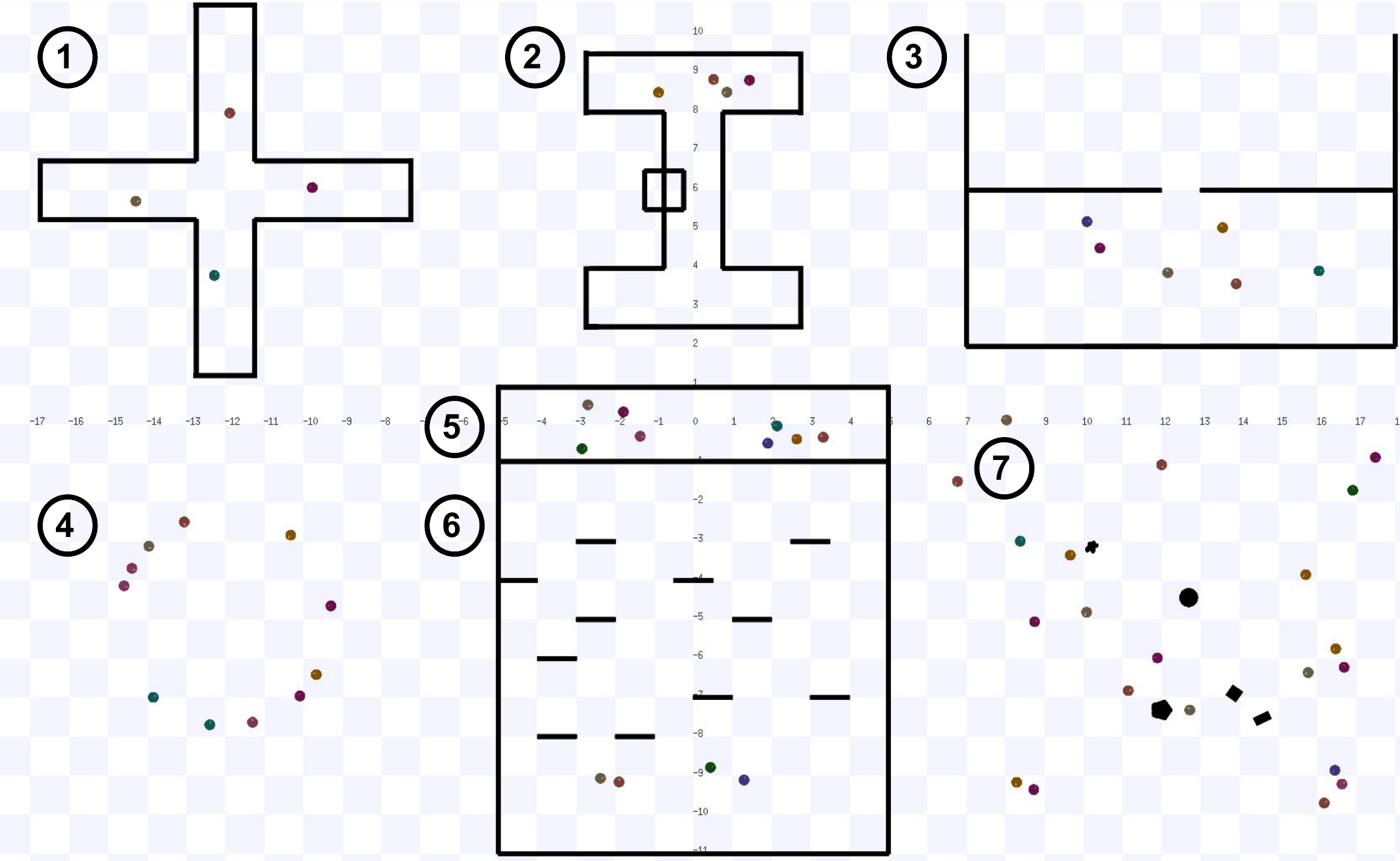}
    \caption{Scenarios corresponding to challenging environments used in the second stage of training. This two-stage training improves the performance.}
    \label{fig:train_scene}
    \vspace{-20px}
\end{figure}

\subsection{Network Training}

Our training strategy extends the method used by learning algorithms based on local information \cite{long2018towards,JHow1}. To accelerate the training process, we divide the overall training computation into two stages. In the first stage, we use $k$ robots (e.g. k=$20$) with random initial positions and random goals in a fully-free environment. In the second stage, we  include a more challenging environments, such as a narrow passage, random obstacles, etc. We show the training scenarios in \pref{fig:train_scene}. 
These varying training environments and the large number of robots in the system can result in a good overall policy. Moreover, the use of global information results in much larger network parameters. We use a $20480\times128$ FC layer, which is more difficult to train than a relatively small, simple network that only accounts for local information. To accelerate the training process and generate accurate results, we do not train the entire network from scratch and instead include pre-training. During the first stage, we retrain the network with additional structures corresponding to the global information (i.e. the global-map) using the pre-trained local information network part. This pre-training stage uses the parameters proposed in \cite{long2018towards}. We highlight the reward as a function of the iterations during the second stage of the training and compare the overall reward computation with that described in the local method \cite{long2018towards} in \pref{fig:reward}. Notice that, since we use a 2D convolutional neural network, our overall training algorithm needs more time during each iteration of the training (ours around 1200s vs~\cite{long2018towards} around 400s). As a result, we do not perform the same number of  training iterations as\cite{long2018towards}, as shown in \pref{fig:reward}. The total training time is around 40 hours.


\section{Implementation and Performance}\label{sec:exp}
In this section, we discuss the performance of our multi-agent navigation algorithm (DeepMNavigate) on complex scenarios and highlight the benefits over prior reinforcement learning methods that only use local information \cite{long2018towards}. 

\begin{figure}[h]
    \centering
    \includegraphics[width=0.3\textwidth]{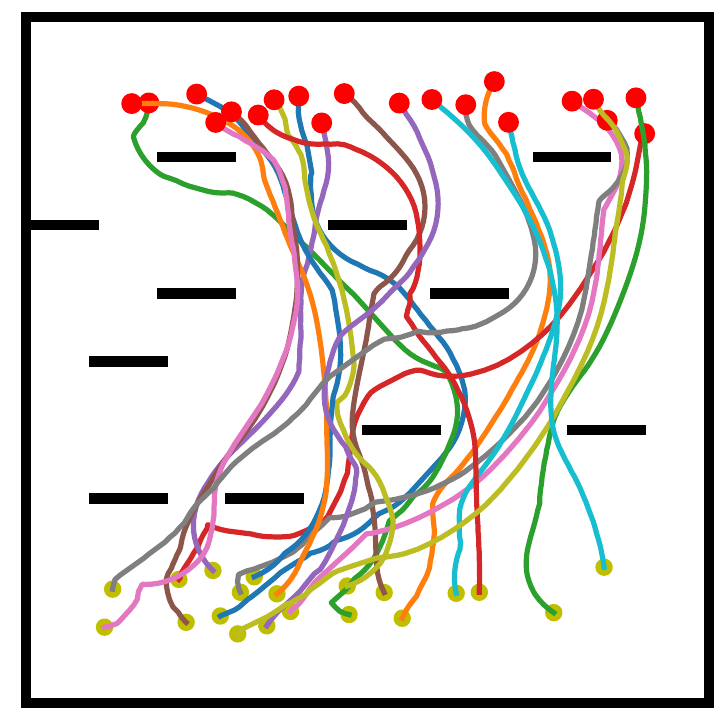}

     \caption{{\bf Room with Obstacles}: We highlight the trajectories of $20$ robots in a room with multiple obstacles. We highlight the initial position of the agent (yellow) and the final position (red) along with multiple obstacles. Prior learning methods that only use local methods~\cite{long2018towards,JHow1} will take more time and may not be able to handle such scenarios when the number of obstacles or the number of agents increases.}
     
    \label{fig:obstacles}
    \vspace{-18px}
\end{figure}

\begin{figure}[h]
    \centering
    \includegraphics[width=0.3\textwidth]{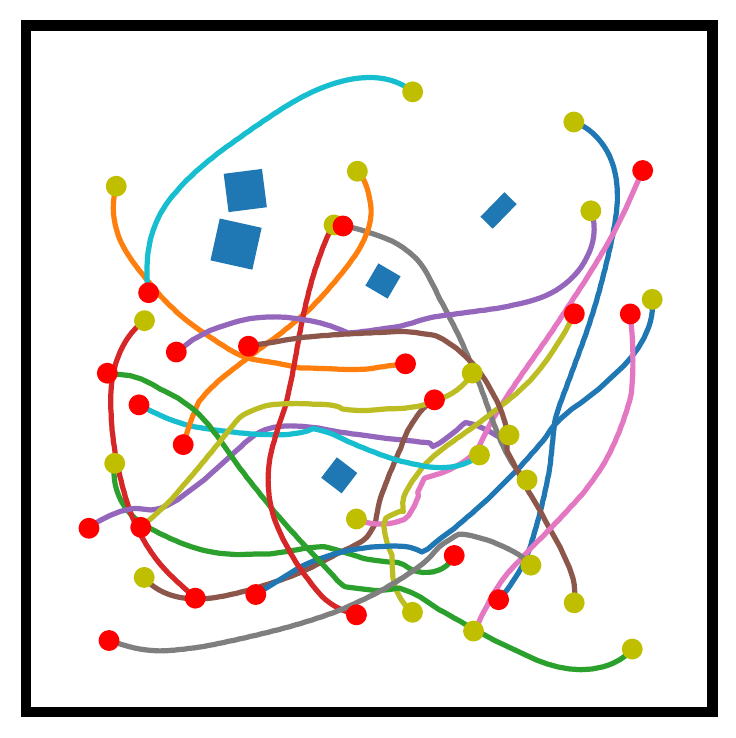}
    \caption{{\bf Random Start and Goal Positions:} Simulated trajectories of $20$ robots from and to random positions in a scene with obstacles. The yellow points are the initial positions and the red points are the final positions. The blue areas are obstacles with random locations and orientations.}
    \label{fig:random_scene}
    \vspace{-15px}
\end{figure}

\subsection{Parameters}

During simulation, the radius is set as $R=0.12m$. In the current implementation, we set $H=W=500m$ for the global-map and $H_l=W_l=250m$ for each local-map. In both situations, we set $w=h=250$. Although a larger map (e.g., $w=h=500$) could include more details from the system, it would significantly increase the network size and the final running time. For instance, the memory requirement of CNN increases quadratically with the input size. In current implementation, we use a PC with 32-core CPU, 32GB memory and one NVIDIA RTX 2080 Ti. To include the additional global information, the algorithm consumes 1.63GB CPU memory, 970MB GPU memory and requires 0.25s for computing one time step, compared to ~\cite{long2018towards} as 1.57 GB CPU memory, 340MB GPU memory and 0.2s. The overhead is not significant. For parameters in the reward function, we set $r_{\arrival} = 15$, $r_{\collision}=-15$, $r_{\text{approaching}}=2.5$, and $r_{\smooth}=-0.1$. We choose $r_{\arrival}$ and $r_{\collision}$ of the same magnitude to obtain an effective, safe behavior, which is a key metric to evaluate robots' trajectories; $r_{\text{approaching}}$ is chosen to encourage robots to approach their goal as fast as they can, which provides a dense feedback to make convergence relatively faster; a small $r_{\smooth}$ value should regularize the trajectory and make it smoother.

\begin{table*}[!htbp]
    \centering
    \footnotesize
    \setlength{\tabcolsep}{4pt}
    \renewcommand{\arraystretch}{1.5}
    \scalebox{0.8}
    {\begin{tabular}{|c|c|c|c|c|c|c|c|c|}
        \hline
        \multirow{2}{*}{Metrics}  & \multirow{2}{*}{Methods} & \multicolumn{7}{c|}{\# of agents (cirle radius (unit:$m$))} \\
        \cline{3-9} 
         &  & 30 (8) & 40 (8) & 50 (8) & 60 (8) & 70 (8) & 80 (12) & 90 (12)\\

        \hline
        \multirow{2}{*}{Success Rate} &  \cite{long2018towards} 
        & $\E{1}$ & $0.975$ & $0.96$ & $0.95$ & $0.929$ & $0.7375$ & $0.722$\\
        \cline{2-9}
         & Ours & $\E{1}$ & $\E{1}$ & $\E{1}$ & $\E{1}$ & $0.986$ & $\E{1}$ & $\E{1} $\\
        
        \hline
        \multirow{2}{*}{Stuck/Collision Rate} &  \cite{long2018towards} 
        & $\E{0/0}$ & $0/0.025$ & $0/0.04$ & $0/0.05$ & $0/0.071$ & $0.175/0.0875$ & $0.233/0.044$ \\
        \cline{2-9}
         & Ours & $\E{0/0}$ & $\E{0/0}$ & $\E{0/0}$ & $\E{0/0}$ & $\E{0/0.014}$ & $\E{0/0}$ & $\E{0/0}$\\

        \hline
        \multirow{2}{*}{Extra Time} & \cite{long2018towards} 
        & $\E{4.32333}$ & $\E{8.20256}$ & $\E{7.8625}$ & $11.3088$ &$13.54$ & $\E{15.1238}$ & $\E{15.3292}$\\
        \cline{2-9}
         & Ours & $8.94667$ & $8.73$ & $9.738$ & $\E{10.1}$& $\E{12.4725}$ & $17.5525$ & $34.1256$\\

        \hline
        \multirow{2}{*}{Average Speed} & \cite{long2018towards} & $\E{0.787272}$ & $\E{0.661087}$ & $\E{0.670508}$ & $0.585892$ & $0.541638$ & $\E{0.54341}$ & $\E{0.610233}$\\
        \cline{2-9}
         & Ours & $0.641368$ & $0.646987$ & $0.621649$ & $\E{0.613027}$ & $\E{0.561946}$ & $0.506294$ & $0.412899$\\
        \hline
    \end{tabular}}

    \vspace*{0.05in}
   
    \scalebox{0.8}{
    \begin{tabular}{|c|c|c|c|c|c|}
        \hline
        \multirow{2}{*}{Metrics}  & \multirow{2}{*}{Methods} & \multicolumn{4}{c|}{\# of agents} \\
        \cline{3-6} 
         &  & 8 & 12 & 16 & 20 \\

        \hline
        \multirow{2}{*}{Success Rate} &  \cite{long2018towards} 
        & $0.875$ & $0.75$ & $0.5625$ & $0.0$ \\
        \cline{2-6}
         & Ours & $\E{1}$ & $\E{1}$ & $\E{1}$ & $\E{1}$\\
         
        \hline
        \multirow{2}{*}{Stuck/Collision Rate} &  \cite{long2018towards} 
        & $0/0.125$ & $0/0.25$ & $0.3125/0.125$ & $1/0$ \\
        \cline{2-6}
         & Ours & $\E{0/0}$ & $\E{0/0}$ & $\E{0/0}$ & $\E{0/0}$\\

        \hline
        \multirow{2}{*}{Extra Time} & \cite{long2018towards} 
        & $4.8$ & $12.5111$ & $30.7222$ & - \\
        \cline{2-6}
         & Ours & $\E{2.7}$ & $\E{4.075}$ & $\E{5.33125}$ & $\E{8.36}$\\

        \hline
        \multirow{2}{*}{Average Speed} & \cite{long2018towards} & $0.653784$ & $0.410699$ & $0.230921$ & - \\
        \cline{2-6}
         & Ours & $\E{0.742279}$ & $\E{0.656969}$ & $\E{0.594551}$ & $\E{0.482519}$ \\
        \hline
    \end{tabular}
    }
    \scalebox{0.8}{
            \begin{tabular}{|c|c|c|c|c|c|}
                \hline
                \multirow{2}{*}{Metrics}  & \multirow{2}{*}{Methods} & \multicolumn{4}{c|}{\# of agents} \\
                \cline{3-6} 
                 &  & 5 & 10 & 15 & 20 \\
        
                \hline
                \multirow{2}{*}{Success Rate} &  \cite{long2018towards} 
                & $\E{1}$ & $0.7$ & $0.6$ & $0.7$ \\
                \cline{2-6}
                 & Ours & $\E{1}$ & $\E{1}$ & $\E{1}$ & $\E{1}$\\
                 
                \hline
                \multirow{2}{*}{Stuck/Collision Rate} &  \cite{long2018towards} 
                & $\E{0/0}$ & $0.1/0.2$ & $0.133/0.267$ & $0.15/0.15$ \\
                \cline{2-6}
                & Ours & $\E{0/0}$ & $\E{0/0}$ & $\E{0/0}$ & $\E{0/0}$\\
        
                \hline
                \multirow{2}{*}{Extra Time} & \cite{long2018towards} 
                & $9.55894$ & $5.13058$ & $6.38476$ & $9.99585$ \\
                \cline{2-6}
                 & Ours & $\E{2.19487}$ & $\E{2.55377}$ & $\E{3.47301}$ & $\E{7.80055}$\\
        
                \hline
                \multirow{2}{*}{Average Speed} & \cite{long2018towards} 
                & $0.707441$ & $0.734304$ & $0.721682$ & $0.508305$ \\
                \cline{2-6}
                 & Ours & $\E{0.858985}$ & $\E{0.828276}$ & $\E{0.775802}$ & $\E{0.582426}$ \\
                \hline
            \end{tabular}
    }
    \caption{The performance of our  method (DeepMNavigate) and prior methods based on local information on different benchmarks (\textbf{Top}: \textit{Circle Crossing}; \textbf{Bottom left}: \textit{Narrow Corridor}; \textbf{Bottom right}: \textit{Room with Obstacles}.), measured in terms of various metrics using different numbers of agents. The bold entries represent the best performance. Our method can guarantee a higher success rate in dense environments as compared to prior multi-agent navigation algorithms.}
    \vspace{-13pt}
    \label{tab:metric}
\end{table*}

\begin{table}[!htbp]
    \centering
    \footnotesize
    \setlength{\tabcolsep}{4pt}
    \renewcommand{\arraystretch}{1.5}
    \scalebox{0.7}{
    \begin{tabular}{|c|c|c|c|c|c|}
        \hline
        \multirow{2}{*}{Metrics}  & \multirow{2}{*}{Methods} & \multicolumn{4}{c|}{\# of agents} \\
        \cline{3-6} 
            &  & 20 & 30 & 40 & 50 \\

        \hline
        \multirow{2}{*}{Success Rate} &  \cite{long2018towards} 
        & $\E{1}$ & $0.867$ & $0.825$ & $0.76$ \\
        \cline{2-6}
            & Ours & $\E{1}$ & $\E{1}$ & $\E{1}$ & $\E{1}$\\
        
        \hline
        \multirow{2}{*}{Stuck/Collision Rate} &  \cite{long2018towards} 
        & $\E{0/0}$ & $0.033/0.1$ & $0.05/0.125$ & $0.2/0.04$ \\
        \cline{2-6}
        & Ours & $\E{0/0}$ & $\E{0/0}$ & $\E{0/0}$ & $\E{0/0}$\\

        \hline
        \multirow{2}{*}{Extra Time} & \cite{long2018towards} 
        & $4.5974$ & $9.38872$ & $13.8348$ & $12.6948$ \\
        \cline{2-6}
            & Ours & $\E{3.59674}$ & $\E{4.06355}$ & $\E{10.4823}$ & $\E{11.2948}$\\

        \hline
        \multirow{2}{*}{Average Speed} & \cite{long2018towards} 
        & $0.601266$ & $0.432238$ & $0.354548$ & $0.332425$ \\
        \cline{2-6}
            & Ours & $\E{0.674891}$ & $\E{0.63543}$ & $\E{0.404211}$ & $\E{0.383049}$ \\
        \hline
    \end{tabular}
    }
    \caption{The performance of our proposed method and prior learning algorithms on the \textit{Random Starts and Goals} benchmark. Our method demonstrates better results (bold face) than prior multi-agent navigation algorithms.}
    \vspace{-10px}
    \label{tab:metric_random}
\end{table}

\begin{figure}[h]
    \centering
    \includegraphics[width=0.3\textwidth]{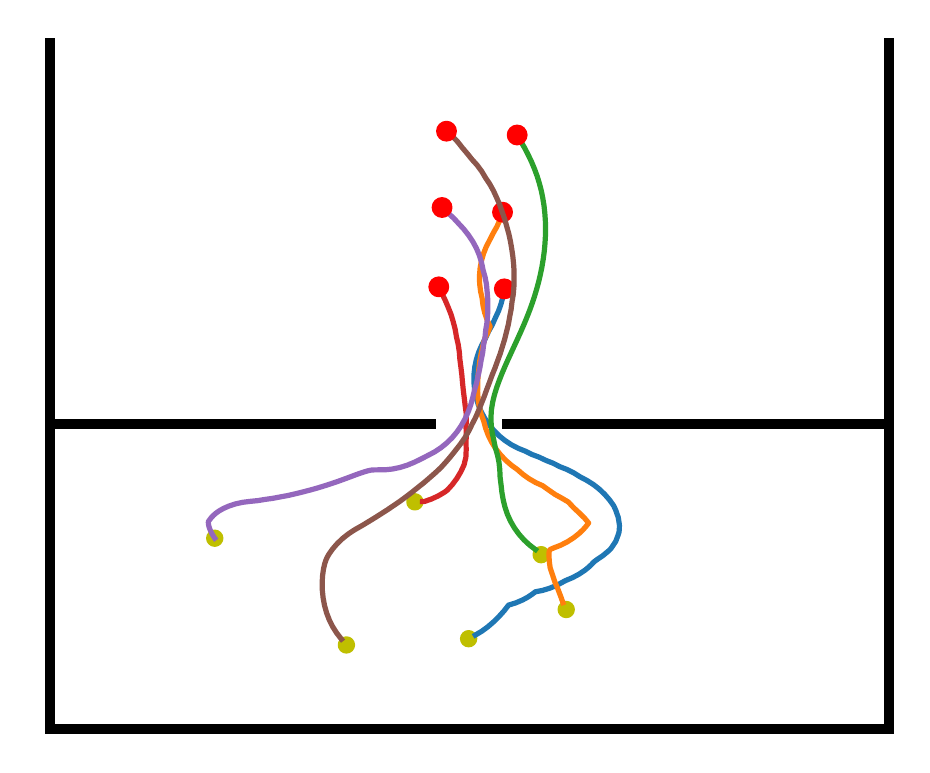}
    \caption{{\bf Room Evacuation:} Simulated trajectories of $6$ robots evacuating a room with our algorithm. The yellow points are the initial positions and the red points are the final positions. Even with a narrow passage, our DeepMNavigate algorithm works well. This benchmark is quite different from training data.}
    \vspace{-8px}
    \label{fig:evacuation}
\end{figure}

\begin{figure}[h]
    \centering
    \includegraphics[width=0.4\textwidth]{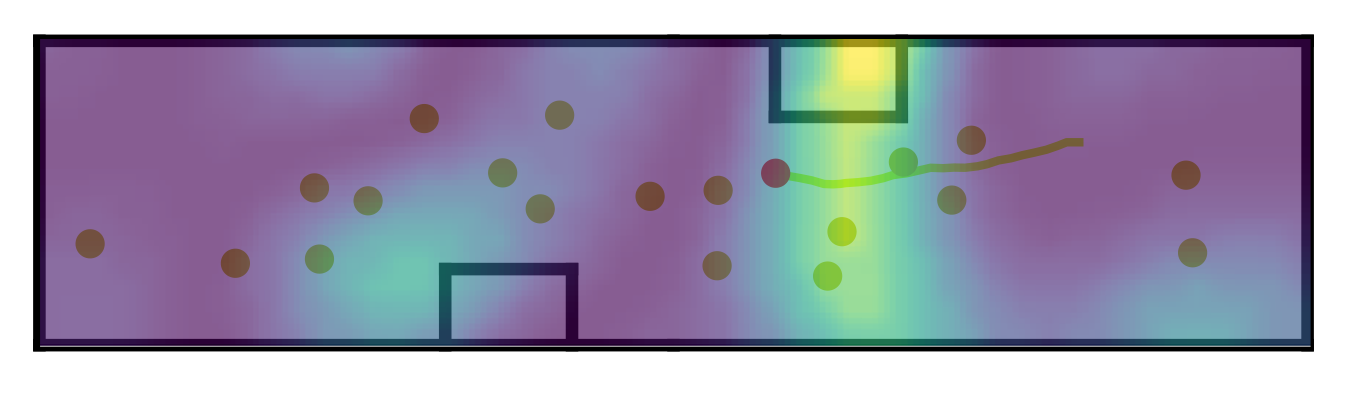}
  \caption{Perturbation saliency generated by method ~\cite{greydanus2017visualizing} in the narrow corridor benchmark. Red point represents the current agent and yellow points represent the surrounding agents. Yellow area is saliency for the action policy.}
    \label{fig:saliency}
    \vspace{-10px}
\end{figure}

\subsection{Evaluation Metrics and Benchmarks}

To evaluate the performance of our navigation algorithm, we use the following metrics:
\begin{itemize}
\item \textit{Success rate}: the ratio of the number of robots reaching their goals in a certain time limit without any collisions to the total number of robots in the environment.
\item \textit{Collision or stuck rate}: if the robot cannot reach the destination within a limited time or collides, they are considered as getting stuck or in-collisions, respectively.
\item \textit{Extra time}: the difference between the average travel time of all robots and the lower bound of the robots' travel time. The latter is computed as the average travel time when going straight towards the goal at the maximum speed without checking for any collisions.
\item \textit{Average speed}: the average speed of all robots during the navigation.
\end{itemize}

We have evaluated our algorithm in five challenging and representative benchmarks:
\begin{itemize}
    \item \textit{Circle Crossing}: The robots are uniformly placed around a circle and the goal position of each robot is on the opposite end of the circle based on the diameter. The scenarios are widely used in prior multi-agent navigation algorithms~\cite{van2011reciprocal,long2018towards} (\pref{fig:traj_circle}).
    \item \textit{Narrow Corridor}: Two groups of robots exchange their positions through a narrow corridor. This benchmark is hard for geometric decentralized methods~\cite{van2011reciprocal}, which cannot navigate robots through narrow passages (\pref{fig:corridor}).
    \item \textit{Room with Obstacles}: The robots cross across a room full of obstacles, from one side to the other. Methods only using local information~\cite{long2018towards} will spend more time finding the path to the goal and may even fail due to lack of global information (\pref{fig:obstacles}).
    \item \textit{Random Starts and Goals}: The robots start from random initial positions and moves to random goal positions. Also, there are obstacles with random locations and orientation. Global information will help find safer and faster paths (\pref{fig:random_scene}).
    \item \textit{Room Evacuation}: The robots start from random initial positions and evacuate to the outside of the room. They need to cross one small door while avoiding collisions (\pref{fig:evacuation}).
\end{itemize}

\subsection{New and Different Benchmarks}
We have evaluated the performance of our method on benchmarks that are quite different from the training data in terms of layout and the inclusion of narrow passages. They also use different numbers of agents. In the Circle Crossing benchmark, we only train with $12$ agents, but evaluate in a similar scene with $90$ agents. Furthermore, some of the benchmarks like Narrow Corridor and Room Evacuation are quite different from the training dataset. DeepMNavigate is still able to compute collision-free and smooth trajectories for all the agents with no collisions and each agent arrives at its goal position. As shown in Fig. \ref{fig:corridor}, the local learning method~\cite{long2018towards}  does not consider the global map information and fails in such scenarios. In contrast, our approach enable robots to learn a reciprocal navigation behavior according to the global map information without any communication on action decision.

\subsection{Quantitative Evaluation}
We have evaluated the performance of our algorithm in terms of different evaluation metrics described above. In the circle crossing scenario, the failure rate of \cite{long2018towards} rises with the increasing number or the density of robots. However, our approach always results in a stable performance and can avoid the deadlock situation. At times, some of the robots may need to take a longer path to avoid congestion and this could reduce the robot's efficiency. 
As a result, we obtain better performance as compared to    \cite{long2018towards} or the decentralized collision avoidance methods in high density benchmarks like Circle Crossing. 

Different from the circle crossing scenario, the other benchmarks  incorporate some static obstacles in the environment. In this case, our method integrates the map information into the policy network and utilizes that information to handle such static obstacles and narrow passages.    As the experimental results shown, our approach outperforms \cite{long2018towards} both in terms of success rate and efficiency metric. In addition, our method behaves robustly even with the increasing density of robots.

Another important criteria to evaluate the performance of multi-robot systems is the stuck or collision rate, which is a measure of the number of robots cannot reach the goals or collide on the road. As shown in Tables II and III, our collision rate is zero for all these benchmarks. On the other hand, techniques based on local navigation information result in some number of failures on different benchmarks. Furthermore, the failure increases as the number of agents or the density increases.

To better understand how the global information help the navigation system, we compute perturbation saliency over global map using method from~\cite{greydanus2017visualizing} in the narrow corridor benchmark. We show the result in \pref{fig:saliency}. The most important areas in the global map to help the decision making include: 1) agents in the front, which are blocked by other agents in local laser scan; 2) agents from the back, which could not be covered by local laser scan; 3) nearby obstacle. Thus, the global information can help the agent plan in advance, and be more alert to nearby obstacles.

\subsection{Scalability}
\begin{figure}[h]
\centering
\begin{minipage}{.45\textwidth}
\centering
\includegraphics[width=1\textwidth]{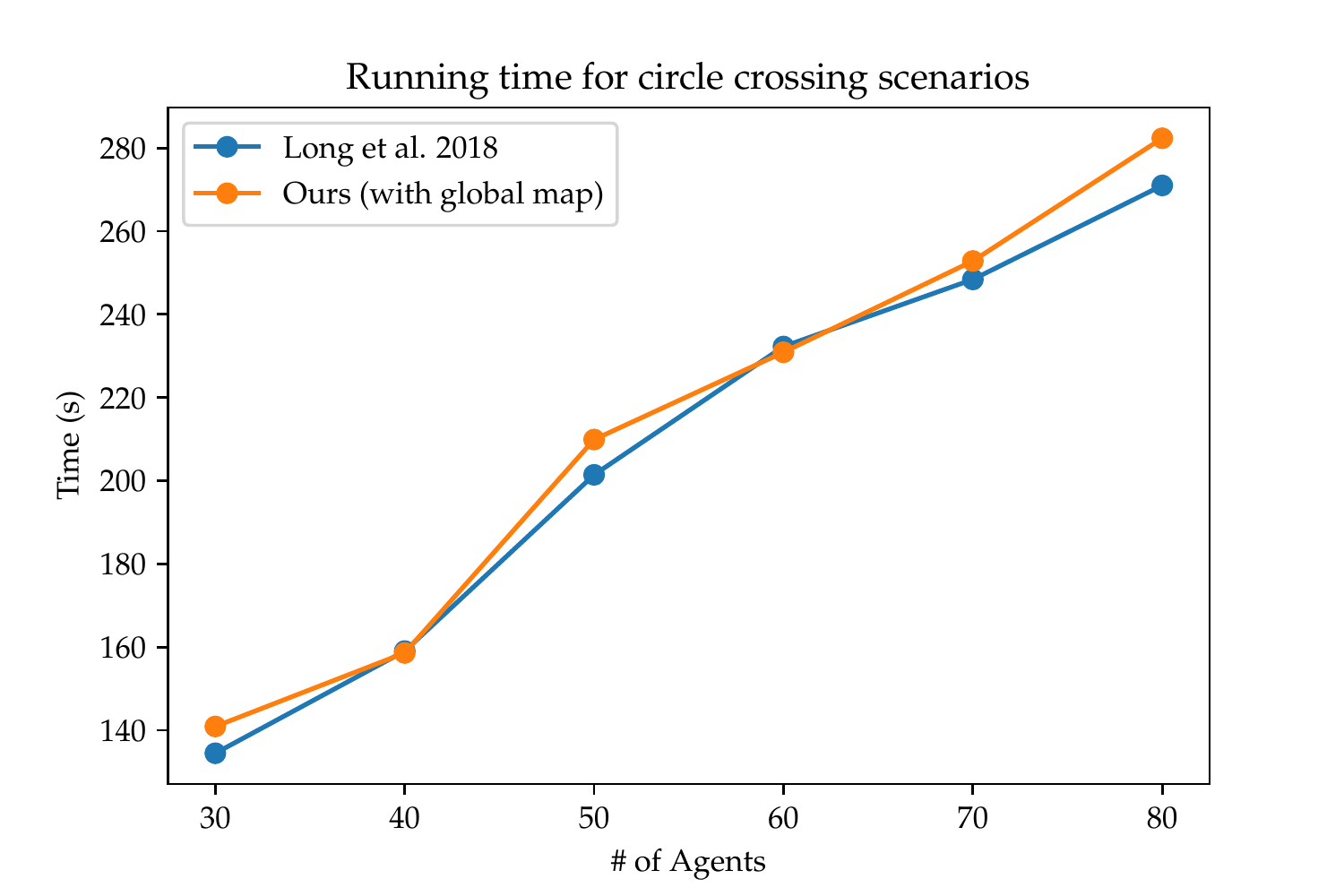}
\end{minipage}
\begin{minipage}{.45\textwidth}
\centering
\includegraphics[width=1\textwidth]{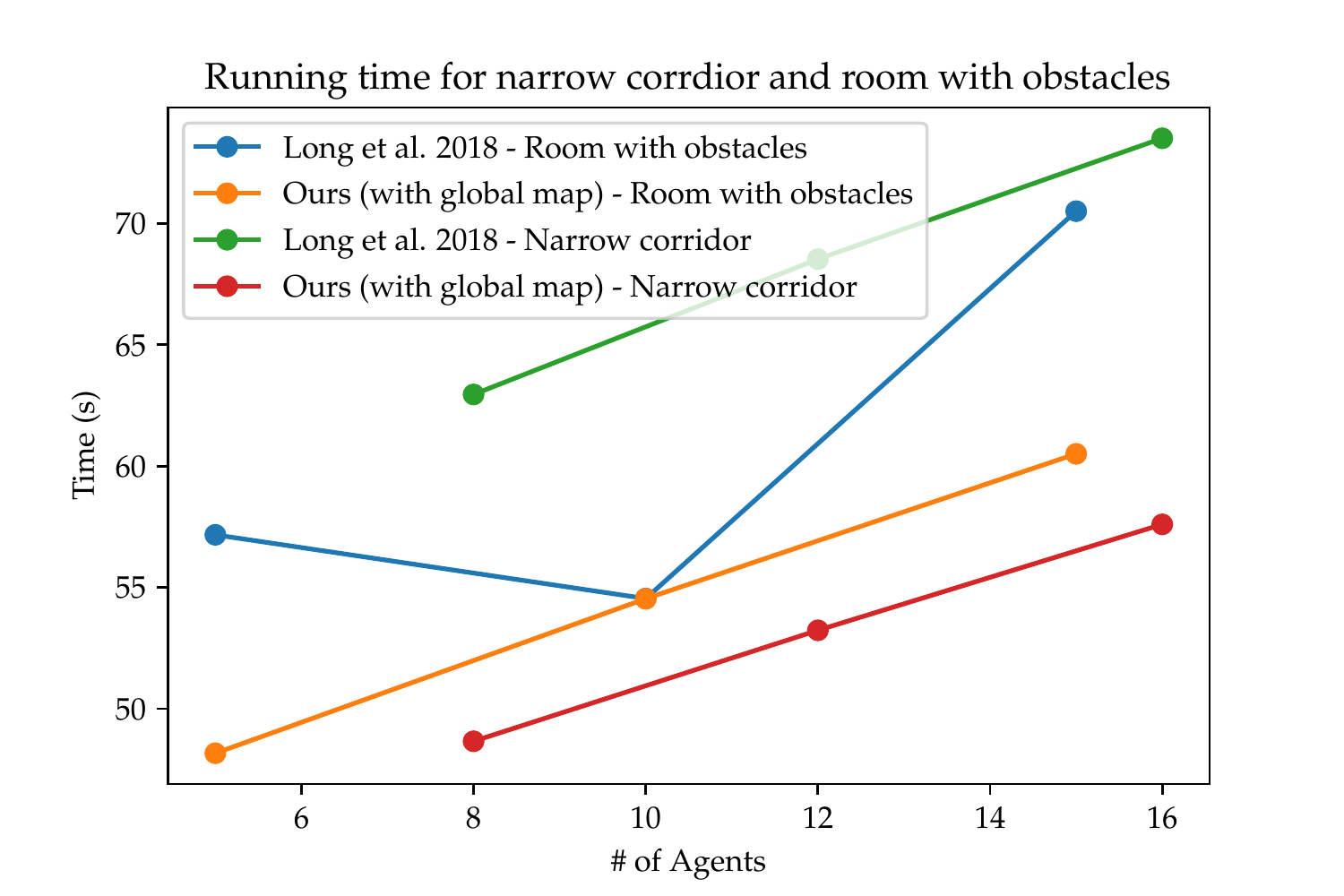}
\end{minipage}
\caption{We show the running time with different numbers of agents in several benchmarks. We use a CPU with 32 cores and NVIDIA RTX 2080 Ti to generate these performance graphs. This timing graph demonstrates that our approach is practical for many tens of robots. We also compare the running time with a learning-based algorithm that only uses local information~\cite{long2018towards}. The additional overhead in the running time with the use of the global information is rather small.}
\label{fig:running_time}
\vspace{-13px}
\end{figure}

The running times for different numbers of robots are shown in \pref{fig:running_time}. We can observe linear time performance with number of agents. That is because the decision process of our method is independent, each agent can compute their action by itself, based on the information it receives. Compared with most traditional geometric-based methods, a step analyzing the global environment may have super-linear time requirements, especially for the congested or challenging benchmarks used in this paper. For example, some methods compute K-nearest neighbors or the roadmap of the environment using the Voronoi diagram, which can have super-linear complexity. In contrast, our approach does not perform any such global computations and only leverages the power of the neural network. Moreover, the computation of the global map takes $O(n)$ time.

\section{Conclusion, Limitations, and Future Work}

We present a novel, multi-agent navigation algorithm based on deep reinforcement learning. We show that global information about the environment can be used based on the global-map and present a novel network architecture to compute a collision-free trajectory for each robot. We highlight its benefits over many challenging scenarios and show benefits over geometric decentralized methods or reinforcement learning methods that only use local information. Moreover, our experimental results show that our DeepMNavigation algorithm can offer improved performance in dense and narrow scenarios, as compared to prior approaches. Furthermore, we demonstrate the performance on new, different benchmarks that are different from training scenarios. Overall, our approach demonstrates the value of global information in terms of discretized maps for DRL-based ethod.

Our results are promising and there are many ways to improve the performance. Our idea can be extended to 3D environments, by replacing the global map with a 3D version and using a 3D convolutional neural network to process it. Current training scenarios do not include dynamic and dense obstacles, so we may include them in the future work. Also, we need better techniques to compute the optimal size of the global-map and the local-map, and we can also include other components of the state information like velocity, orientation, or dynamics constraints. We also need to extend the approach to handle general dynamic scenes where no information is available about the motion of the moving obstacles. Our approach assumes that the global information is available and that, in many scenarios, obtaining such information could be expensive. 
The use of global information increases the complexity of the training computation and we use a two-stage algorithm to reduce its running time. One other possibility is to use an auto-encoder to automatically derive the  low-dimensional feature representations and then use them as a feature extractor.  It may be possible to split the global and local navigation computation to combine our approach with local methods to avoid collisions with other agents or dynamic obstacles \cite{van2011reciprocal,geraerts2008using}. Such a combination of local and global methods has been used to simulate large crowds~\cite{narain2009aggregate} and it may be useful to develop a similar framework for learning-based algorithms. Furthermore, DRL-based methods that use local or global information can also be combined with global navigation data structures (e.g., roadmaps) to further improve navigation performance.

We have only demonstrated the application of our DRL-method on challenging, synthetic environments. A good area for future work is extending to real-world scenes, where we need to use other techniques to generate the motion information maps. It would be useful to combine our learning methods with SLAM techniques to improve the navigation capabilities.

\section{Acknowledgement}
This work was supported by ARO grants
(W911NF1810313 and W911NF1910315) and Intel.
Tingxiang Fan and Jia Pan were partially supported by HKSAR General Research Fund (GRF) HKU 11202119, 11207818.

\small
\bibliographystyle{IEEEtran}
\bibliography{bibliography}
\end{document}